\documentclass[twocolumn,tighten]{aastex62}
\usepackage{amssymb,amsmath}
\usepackage{bm}% bold math
\usepackage{mathtools} %include \prescript
\usepackage{hyperref}% add hypertext capabilities
\newcommand{\rhat}{\hat{\bm r}}
\newcommand{\rihat}{\hat{\bm r}_i}
\newcommand{\rjhat}{\hat{\bm r}_j}
\newcommand{\pij}{\langle ij \rangle}
\begin{document}

%\preprint{APS/123-QED}

\title{ Pairwise \emph{Transverse} Velocity Measurement with the Rees-Sciama Effect}% Force line breaks with \\}%

\author{Siavash Yasini}
\email{yasini@usc.edu} 
\author{Nareg Mirzatuny}
\email{mirzatun@usc.edu}
\author{Elena Pierpaoli}
\email{pierpaol@usc.edu}
%
%\altaffiliation[Also at ]{}%Lines break automatically or can be forced with \\
% \email{}
\affiliation{%
 Physics \& Astronomy Department, University of Southern California, Los Angeles, California,  90089-0484 \\
}%

\date{\today}% It is always \today, today,

\begin{abstract}
We introduce a new estimator for the mean pairwise velocities of galaxy clusters, which is based on the measurement of the clusters' $\textit{transverse}$ velocity components. The Rees-Sciama (RS) effect offers an opportunity to measure transverse peculiar velocities through its distinct dipolar signature around the halo centers in the Cosmic Microwave Background (CMB) temperature map. We exploit this dipolar structure to extract the magnitude and direction of the transverse velocity vectors from CMB maps simulated with the expected characteristics of future surveys like CMB-S4. Although in the presence of lensed CMB and instrumental noise individual velocities are not reliably reconstructed, we demonstrate that the mean pairwise velocity measurement obtained using the estimator yields a signal-to-noise ratio of $5.2$ for $\sim21,000$ halos with $M > 7\times10^{13}\rm M_\odot$ in a $40\times40$ [deg$^2$] patch at $z=0.5$. While the proposed estimator carries promising prospects for measuring pairwise velocities through the RS effect in CMB stage IV experiments, its applications extend to any other potential probe of transverse velocities.

\end{abstract}

%Use showkeys class option if keyword
%\maketitle

%\tableofcontents

\section{Introduction}\label{sec:intro}
The study of the peculiar velocity field offers an alternative way to constrain cosmology other than the inspection  of density anisotropies. 
All concrete ways to characterize peculiar velocities to date are based on the measurement of the \emph{radial} components. This is obtained  either through the inspection of galaxies' spectra, or via the study of the kinetic Sunyaev-Zeldovich (kSZ) effect \citep{Sunyaev:1972eq} in CMB maps (see, e.g. \cite{Mroczkowski:2018nrv, Roncarelli:2017cwe,Roncarelli:2018kud, Bhattacharya:2007sk,Hill:2016dta,Mak:2011sw}).
The amplitude of the kSZ is $\sim 1 \mu$K and hence subdominant to the primary CMB fluctuations at $\ell \lesssim 4000$, however, it is possible to increase its  signal-to-noise ratio (S/N) using a differential (pairwise) measurement; averaging the pairwise kSZ signal over several pairs naturally produces a cancellation of noise sources. The first detection of pairwise velocities \citep{Hand:2012ui} was achieved using an   estimator for the signal \citep{Ferreira:1998id} which employed line-of-sight (LOS) measurements of peculiar velocities. Although LOS  velocities, under the assumption of large scale isotropy, well represent the underlying 3D field, the remaining 2/3 of the information is contained in the yet--undetected transverse components.

Measuring transverse velocities through investigation of spectroscopic redshifts is certainly challenging because the Doppler shift they generate is  second order in velocity $v/c$ \citep{Zhao:2012st,Wojtak:2011ia} and their extraction would also require an independent measurement of the LOS velocities. It is not plausible to measure individual transverse velocities through angular displacement either:  a halo at a comoving distance of $\sim1$ Gpc $(z\sim0.2)$ with a peculiar velocity of 1000 km/s moves at a rate of 1 arcsec per 10 million years. However, it has been shown that using correlation functions can lead to a statistical detection of transverse velocities for low redshift objects  \citep{Hall:2018uug, Darling:2018jxu}.
%(in particular using correlation functions) can help overcome this chall

Alternative approaches have been proposed to detect transverse velocities such as observing the frequency shifts of a background radiation with sharp features in multiple images of a strong lens \citep{Molnar:2002pf}. This method, however, requires extremely high S/N in the spectral measurement. As for  CMB measurements the transverse velocities can be probed by: 

(i)~The polarized kSZ effect \citep{Sunyaev:1980nv,Sazonov1999,Yasini:2016pby}. This measurement requires sensitivity levels around $\sim100$~nK (depending on the velocity and optical depth) and $\sim$1 arcminute resolution (depending on angular size and redshift). Given the specifications of proposed CMB surveys \citep{CMB-S4tech,CMB-S4science,SimonsObservatory}, this effect has a promising prospect for detection in the foreseeable future. However, it has an inherent parity degeneracy regarding the direction of the velocity vector and it is  only sensitive to the orientation \citep{Sazonov1999}. 

(ii)~The Rees-Sciama effect (RS) \citep{Rees:1968zza} induced by the transverse motion of clusters. This is  also known as the Moving Cluster of Galaxies (MCG) \citep{Molnar:2000de} or Birkinshaw \& Gull (BG) \citep{birkinshaw1983test} or the Moving Lens effect. 
The RS effect induces a dipolar temperature fluctuation around the center of the halo in the plane of the sky \citep{Aghanim:1998ux,Cai:2010hx,Tuluie:1995ut,Tuluie:1995pg}, in the direction of the cluster's transverse motion.
Having the largest amplitude among the aforementioned effects, the RS effect seems the most promising way to detect transverse velocities, and it is therefore the focus of this work.

In this paper, we introduce a mean pairwise velocity estimator which employs the transverse components. The transverse velocities are reconstructed through the RS effect in simulated CMB maps using a heuristic filter. We then apply the estimator to the reconstructed velocities and  evaluate  the detection of pairwise velocities in future CMB stage IV experiments.
We show that, even without an optimal matched filter for individual velocity reconstruction, application of the proposed estimator yields a significant S/N for CMB-S4-like experiments. 

In \S\ref{sec:transverse_vel} we derive the the mean pairwise velocity estimator. Filter and template design are described in \S\ref{sec:filter} and their applications to the simulated maps (\S\ref{sec:simul}) are described in \S\ref{sec:analysis}. Overview of the results with a summary and discussion are presented in \S\ref{sec:concl}.

In what follows, $c$, $h$ and $T_z$ are respectively the speed of light, the Hubble factor and the CMB temperature at redshift $z$. In producing the CMB mock maps we use a $\Lambda$CDM cosmology with $(h, \omega_b, \omega_c,\Sigma m_\nu / \textrm{eV},\tau , A_s , n_s) = (67.7, 0.022, 0.119, 0.060, 0.066, 2.30\times10^{-9}, 0.967)$.

\section{Transverse Velocity}\label{sec:transverse_vel}
    \subsection{Pairwise Estimator}\label{subsec:pw_estimator}
    
\begin{figure}[t]
		\centering
		\includegraphics[width=0.55\columnwidth]{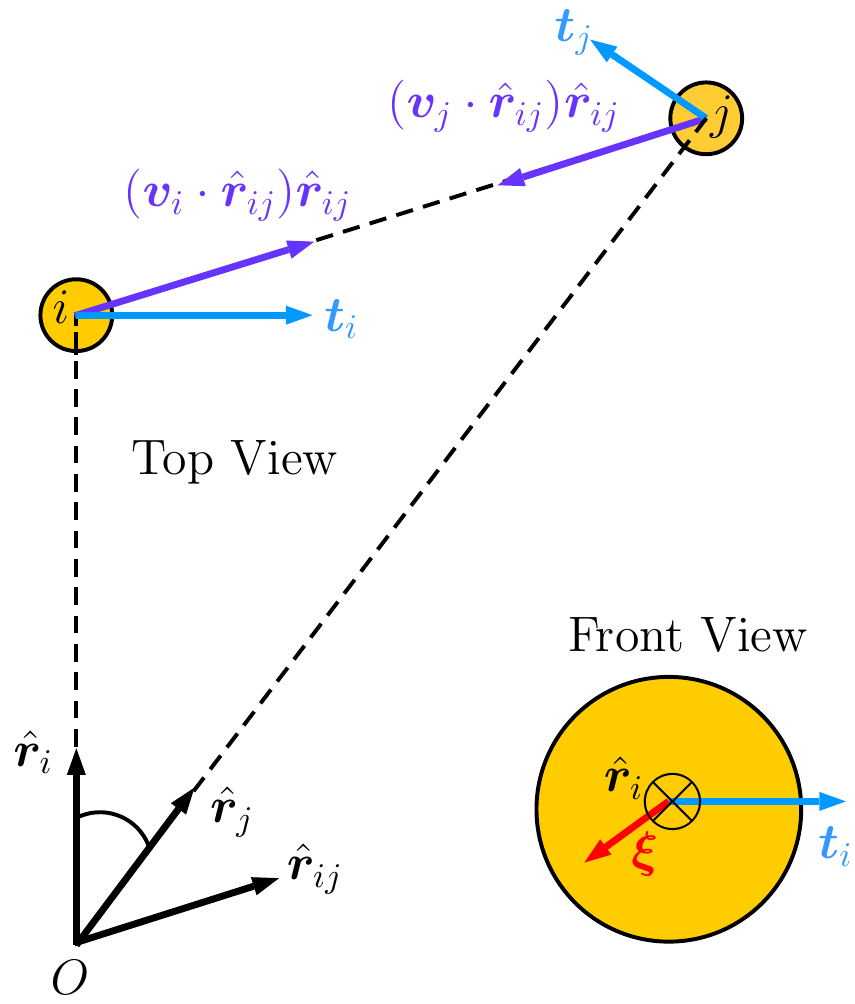}
		\caption{Geometric configuration of the problem.}
		\label{fig:clusters}
	\end{figure}

	The first estimator for the pairwise velocity using the LOS component was presented in \cite{Ferreira:1998id}. In this section we introduce a similar estimator which uses the transverse component instead. We denote the mean pairwise velocity between all the pairs $\{i,j\}$ that are at a distance $r$  from each other as
	
	\begin{equation}
		v_{\pij}(r) \equiv \langle \bm{v}_{ij}(r) \cdot \rhat_{ij} \rangle = \langle (\bm{v}_{i} -\bm{v}_{j}) \cdot \rhat_{ij} \rangle
	\end{equation}
	where $\bm r_{ij} = r \rhat_{ij} \equiv \bm{r}_j - \bm{r}_i $ for a pair of clusters at locations $\bm{r}_i$ and $\bm{r}_j$, with peculiar velocity vectors $\bm{v}_i$ and $\bm{v}_j$ (see Fig.~\ref{fig:clusters}). Here the subscript $\langle ij \rangle$ symbolically represents the pair-weighted average over the samples and does not have any tensorial meaning. We use this more expressive notation instead of the traditional subscript 12 ($v_{12}$). 
%	We find this notation more expressive than the traditional notation, $v_{12}$.
	
	%For a pair of clusters at locations $\bm{r}_i$ and $\bm{r}_j$, the individual pairwise velocity vector on average points along $\bm{r}_{ij} \equiv \bm{r}_i - \bm{r}_j$, so we have $\bm{v}_{\pij}=v_{\pij} \hat{\bm r}$.  Assuming that the velocity vectors of clusters statistically follow the pairwise velocity vector, we can write

	%\begin{subequations}\label{vi_vj}
	%	\begin{align}
	%		\bm{v}_{i} &= \frac{{v}_{\pij}}{2}  \hat{\bm r} , \\
	%		\bm{v}_{j} &= -\frac{{v}_{\pij}}{2}  \hat{\bm r} .
	%	\end{align}
	%\end{subequations}

	%We emphasize that these relations hold in a statistical sense, meaning that the components perpendicular to $\rhat$ are assumed to be distributed randomly and therefore vanish in any averaging process. 
	
	It is possible to estimate the pairwise velocity using the transverse components 
	\begin{subequations}
		\begin{align}
		\bm{t}_{i} &= \rihat \times (\bm{v}_{i} \times \rihat) , \\
		\bm{t}_{j} &= \rjhat \times (\bm{v}_{j} \times \rjhat) .
		\end{align} 
	\end{subequations}
	 In spherical coordinates the transverse velocities are 2D vectors which can be decomposed as $\bm t_i = t^\theta_i \hat{\bm \theta}_i + t^\phi_i \hat{\bm \phi}_i $, where $\hat{\bm \theta}_i = \partial \rihat/\partial \theta_i$ and $\hat{\bm \phi}_i =  \partial \rihat/\partial \phi_i (\sin\theta_i)^{-1}$.
	 
	  The pairwise transverse velocity for a pair of objects $i$ and $j$ can be written as
	\begin{align}\label{ti-tj}
		\bm{t}_{i} - \bm{t}_j &=  \rihat \times (\bm{v}_{i} \times \rihat) - \rjhat \times (\bm{v}_{j} \times \rjhat).
	\end{align}
	Taking the average of both sides for all the pairs at a distance $r$ from each other yields
	\begin{align}\label{ti-tj2}
		\langle \bm{t}_{i} - \bm{t}_j \rangle &= 
		v_{\pij}(r)\left\langle [2\rhat_{ij} - \rihat(\rhat_{ij} \cdot \rihat) - \rjhat(\rhat_{ij} \cdot \rjhat) ]/2\right\rangle \nonumber\\
		&= v_{\pij}(r) ~\bm{q}_{\pij},
	\end{align}
	where 
	\begin{equation}
		 \bm{q}_{ij} \equiv [2\rhat_{ij} - \rihat(\rhat_{ij} \cdot \rihat) - \rjhat(\rhat_{ij} \cdot \rjhat)]/2 .
	\end{equation}
	Herer we have assumed that the only non-vanishing components of the peculiar velocities in the average are along $\rhat_{ij}$, and that $\bm{v}_{ij}(r) \cdot \rhat_{ij}$ and $\bm{q}_{ij}$ are uncorrelated.
	
	Minimizing  $\chi^2 =\sum_{ij}|(\bm{t}_{i} - \bm{t}_j)-v_{\pij} ~\bm q_{ij}|^2$  yields the estimator for $v_{\pij}$ in terms of transverse velocities as

	\begin{equation}\label{eq:vij_t}
	\boxed{
		\tilde{v}_{\pij}(r) = \frac{\sum_{ij} (\bm t_i -\bm t_j ) \cdot \bm q_{ij} }{\sum_{ij} |\bm q_{ij}|^2}
	}
	\end{equation}
	where tilde denotes the statistical estimation and the sum is over all pairs $i$ and $j$ that are at a distance $r$ from each other. %Fig.~\ref{fig:vij} shows the direct evaluation of the pairwise velocities calculated from the  3D simulation and compares it with the results obtained from the estimator in Eq.~\eqref{eq:vij_t}. For small separations the two lines are indistinguishable. 

	\subsection{Rees-Sciama Effect}\label{sec:RS}
	The transverse velocity of galaxy clusters can be probed via the RS effect induced by the peculiar motion of the cluster. The expected  change in the CMB temperature, for photons passing through halo $i$ at redshift $z$ with a bulk transverse velocity $\bm t_i$ is \citep{birkinshaw1983test,gurvits1986,RubinoMartin:2004pe,Cooray:2005my,aso2002,itoh2009method, Meerburg:2017xga}

	\begin{equation}\label{eq:T_RS}
	    \Theta_i(\bm\xi) = - \frac{\bm{t_i}}{c} \cdot \bm{\beta_i(\bm\xi)},
	\end{equation}
	where $\Theta = \Delta T / T_z$, $\bm \beta$ is the lensing deflection vector (See section 11.4 in \cite{Lewis:2006fu}) and $\bm\xi$ is a 2D vector in the plane of the sky pointing from the center of the halo ($\rihat$) to each observed pixel (see Fig.~\ref{fig:clusters}).

 	\begin{figure}
 	    \centering
 	    \includegraphics[width=\columnwidth]{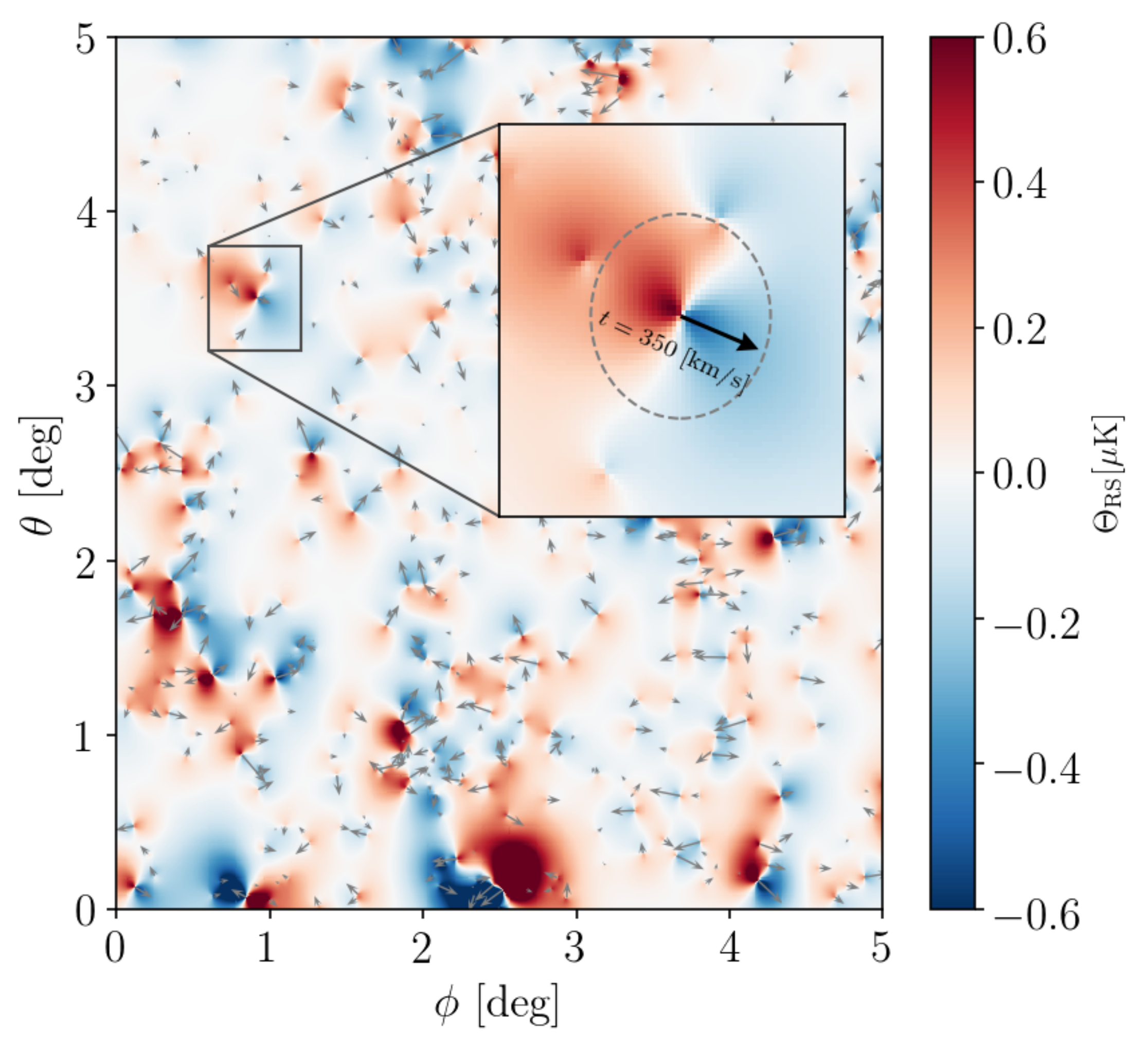}
 	    \caption{Rees-Sciama (RS) effect simulated for clusters with NFW profiles. The grey arrows located at the center of each halo represent the transverse velocity vectors. The zoom-in panel shows the effect for an individual halo of mass 4$\times10^{14} M_\odot$ moving with the transverse velocity $|\bm t|~=~350~$km/s  in the direction indicated by the black arrow. The peak amplitude of the RS effect for such a halo is about 0.6 $\mu$K. The radius of the dashed circle is 3 times the virial radius of the aforementioned halo. }
 	    \label{fig:T_RS_map}
 	\end{figure}
 	
 	%\SY{Maybe remove the first line or change it to: Fig 2 shows the spatial shape of the effect and it forms a dipole around the halo center.}
 	Fig.~\ref{fig:T_RS_map} represents a portion of the simulated map used in the analysis in which the clusters are modeled with an NFW profile. The zoom-in panel explicitly shows the dipolar pattern induced by the RS effect around a particular cluster (Eq.~\ref{eq:T_RS}).
 	%In Eq.~\eqref{eq:T_RS}, the dot product between $\bm t_i$ (a constant vector in the plane of sight) and $\bm \beta$ (pointing radially outwards from the center of halo) causes the RS effect to create a dipolar temperature pattern around the halo (see the zoom-in panel in Fig.~\ref{fig:T_RS_map}). 
 	This pattern can be understood intuitively as a redshift/blueshift of photons as they traverse the moving overdensity: the photons passing through the gravitational field ahead of the cluster ($\bm{t_i} \cdot \bm{\beta_i} >0$) enter at a shallower potential than they exit, and therefore lose energy. The opposite happens in the wake of the cluster ($\bm{t_i} \cdot \bm{\beta_i} <0$) and the photons gain energy upon departure. 
 	The overall result is  a dipolar temperature pattern around the cluster. 
 	
 	Since RS is a purely gravitational effect, it does not depend on the baryonic physics of the clusters, and is only a function of their mass profile. It is evident from Fig.~\ref{fig:T_RS_map} that the RS effect extends way beyond the extent of the halos. As the referee pointed out, this is because the coherence lengths of the transverse velocity and the potential gradient are both larger than the virial radius of the halos. This causes the signals from nearby clusters to overlap, leading to either  cancellation or amplification of the effect integrated along the line of sight. Although present in the simulated maps, we do not model this complication in our filtering process or analysis. Additionally, since the observed RS effect is integrated along the line of sight, the signals from clusters at other redshift bins can also overlap. However, since the spatial filter that we use to extract the signal (see next section) mostly relies on the pixels at a close proximity to the cluster's center (half the virial radius) we do not expect a large number of these overlapping events and therefore ignore their associated error in the analysis. %Also, since the width of our redshift bin is smaller than the correlation length 

\section{Differential Gaussian Derivative Filter}\label{sec:filter}

    We design a heuristic filter which takes advantage of the signal's dipolar shape in order to extract the amplitude and direction of the transverse velocity. Applying a spatial filter $\bm{\Psi}_i(\bm\xi)$ to both sides of Eq.~\eqref{eq:T_RS} over a patch around the cluster $i$ yields
    
    \begin{equation}\label{eq:filter_general}
        \int_i \Theta_i (\bm\xi) \bm\Psi_i(\bm\xi) \mathrm{d^2} \bm\xi = \frac{-1}{c} \int_i \bm t_i \cdot \bm{\beta}_i(\bm\xi)\bm\Psi_i(\bm\xi)  \mathrm{d^2}\bm\xi.
    \end{equation}
     Notice that $\bm t_i$ is assumed to be a constant vector over the extent of the halo and it does not have a $\bm\xi$ dependence. Due to the anti-symmetric nature of the signal, it is obvious that anti-symmetric filters must be used to avoid a vanishing integral on the right hand side. We use $\bm \Psi_i(\bm \xi) = \Psi^\theta_i(\bm \xi) \hat{\bm \theta}_i + \Psi^\phi_i (\bm \xi)\hat{\bm \phi}_i$, where each component is an anti-symmetric (odd function)  filter along the indicated coordinate axis. Assuming azimuthal symmetry, by fixing the filtering axis to $\hat{\bm \theta}_i$ and factoring out the transverse velocity in Eq.~\eqref{eq:filter_general}, we obtain

    \begin{equation}\label{eq:filtered_t_i}
         t^\theta_i  = 
        \frac{-c\int_i \Theta_i (\bm\xi)~ \Psi^\theta_i(\bm\xi)~ \mathrm{d^2} \bm\xi}
             {\int_i \hat{\bm\theta}_i \cdot \bm{\beta}_i(\bm\xi) ~ \Psi^\theta_i(\bm\xi)~  \mathrm{d^2}\bm\xi}.
    \end{equation}
    By replacing $\hat{\bm \theta_i}$ with $\hat{\bm \phi}_i$, we obtain the expression for the reconstructed $t^\phi_i$ component.

    In order to extract the direction of the transverse velocity (or the signs of $t^\theta_i$ and $t^\phi_i$) we need a dipolar filter to match the shape of the signal; if the two are aligned (anti-aligned) we get a positive (negative) number, which indicates the direction. The most immediate and straightforward choice for such a dipolar filter is a Gaussian derivative (GD) filter which is typically used for edge detection in image processing. 
    %A quick glance at Fig.~\ref{fig:T_RS_map} reveals how the velocity field creates white ``edges" in the temperature map separating the hot and cold spots, justifying our heuristic choice of this class of filters. 
    
\begin{figure}[t!]
        \centering
        \includegraphics[width=\columnwidth]{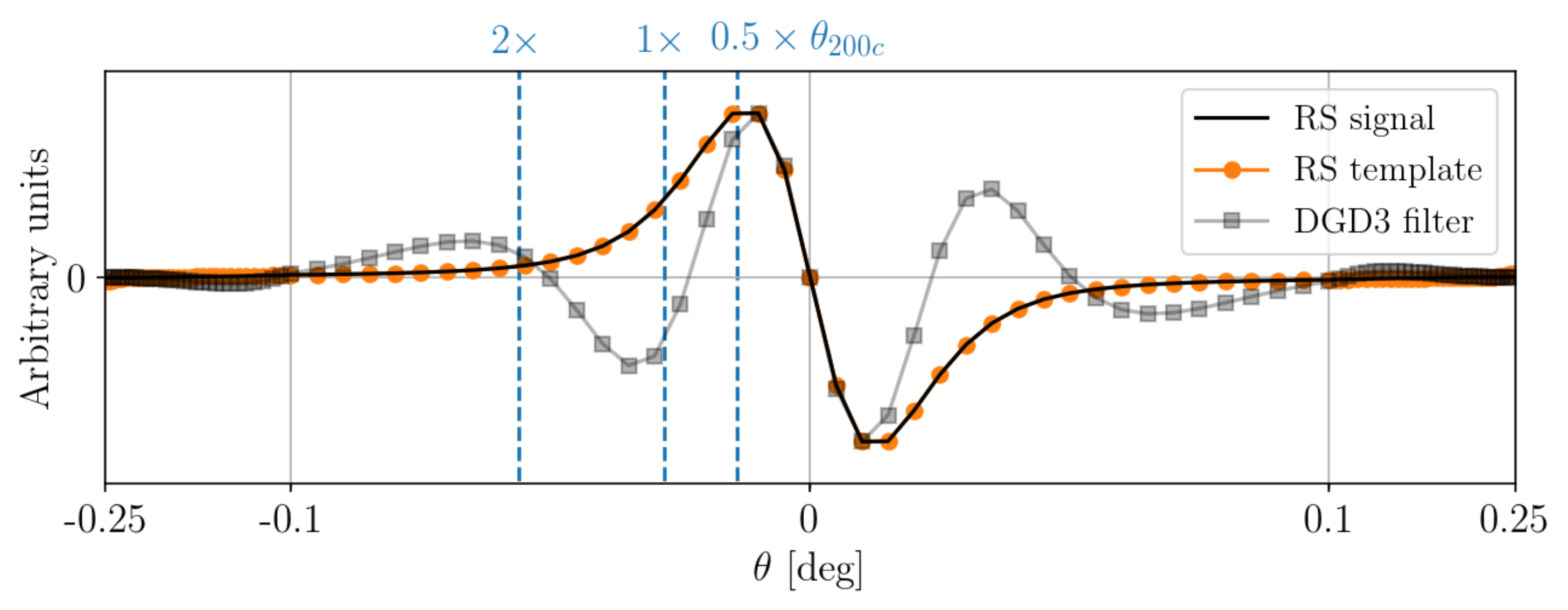}
        \caption{1D profile of the RS signal, the RS template and the differential Gaussian derivative of 3rd order (DGD3) filter (for an NFW profile).  Both the signal and the template have been subjected to a spatial Gaussian high pass filter (FWHM=$5'$) and then smoothed with a 1 arcmin beam. Here since the halo is moving towards the positive $\theta$-axis, the template and the signal are identical up to a normalization factor ($t_i^\theta$) . The circles and squares indicate the function values of the template and the filter at each pixel. The scale of the plot on the $\theta$ axis between $0$ to $0.1$ is linear, and logarithmic afterwards. The dashed blue lines indicate various length scales in terms of the angular radius of the halo, $\theta_{200c}$.}
        \label{fig:profiles}
    \end{figure}

    We represent the n-th derivative of a Gaussian as $\bm\nabla^n \mathcal G_\sigma(\bm\xi)$ where $\bm\nabla$ is a 2D derivative with respect to $\bm \xi$ and  
    \begin{equation}
        \mathcal{G}_\sigma (\bm\xi) \equiv \frac{e^{-|\bm\xi|^2/(2\sigma^2)}}{2\pi\sigma^2}.
    \end{equation}
    By stacking three derivatives with alternating signs, we build the following filter 
    \begin{equation}\label{eq:DGD3_filter}
        \bm \Psi^{\mathrm{DGD3}}_\sigma \equiv 
        \bm\nabla \bm\nabla^2  [\mathcal G_{\sigma/2}-
                \mathcal G_\sigma +
                \mathcal G_{2\sigma}](\bm\xi),
    \end{equation}
    which we call the Differential Gaussian Derivative of 3rd order (DGD3). We set $\sigma$ to be the angular projection of the halo virial radius $\theta_{200c}$. DGD3 can be thought of as a dipolar generalization of the aperture photometry filter used for kSZ detection \citep{Li:2017uin,Soergel:2016mce,DeBernardis:2016pdv,Schaan:2015uaa}. Evaluating Eq. \eqref{eq:DGD3_filter} yields an expression that is proportional to $\bm \xi$, satisfying the anti-symmetric property of the filter. The 1D profile of the filter along its anti-symmetric axis is shown in Fig.~\ref{fig:profiles}. 
    
    If we were to extract the transverse velocities from an RS-only map, a simple Gaussian derivative ($\bm\nabla \mathcal G_\sigma$) would have sufficed. However, in the presence of coherent noise components (e.g. primary CMB fluctuations) any other dipolar temperature gradients would also be picked up as spurious signal by such a filter, and therefore   compromise the velocity reconstruction process. In order to mitigate this risk, we stack three Gaussian derivatives which add alternating side lobes to the tail of the filter. This way the filter eliminates any residual gradient that  extends beyond its first peak. In order achieve this goal, we  push the side lobes far enough from the center, and we use the 3rd derivative ($n=3$) of the Gaussian. We enhance the performance of the filter further by adding a pre-filtering step to the process. Before we extract the velocities with the DGD3 filter, we perform a spatial Gaussian high pass filter (HPF) with FWHM=5 arcminutes to suppress any large gradients in the map. We also apply the HPF to the NFW spatial template ($\hat{\bm\theta}_i \cdot \bm{\beta}_i(\bm\xi)$ ) that we employ for the normalization of the filter (denominator of Eq.~\eqref{eq:filtered_t_i}; see RS template in Fig.~\ref{fig:profiles}).

\section{Simulations and Mapmaking}\label{sec:simul}

	We use the MICE N-body simulation \citep{Crocce:2013vda} at $z=0.5$ to create a mock map of the RS  effect using the locations, masses and peculiar velocities provided by the catalog. The simulation consists of $2048^3$ dark matter particles with a particle mass of $\sim 23 \times 10^{10} \textrm{M}_{\odot}/h$ in a box with $L=3072~ \textrm{Mpc}/h$. 
     We cut a $1000^2\times600~\rm Mpc^3$ slab ($\Delta z \sim 0.2$) from the center of the box and place it at a comoving distance of 1900 Mpc away from the observer. We neglect any time evolution across the slab and assume that all the halos are at the same redshift. The size of the final map that we acquire is about $40 \times 40~\rm deg^2$ with a pixel size of 0.3 arcmins.
	
	After applying a mass-cut of $7\times10^{13}\rm M_\odot$ (chosen merely for numerical expediency), we are left with about 21,000 halos in the catalog. We assign NFW density profiles to the halos and analytically calculate the lensing deflection angle using the formulas in \citep{Baxter:2014frs,Mood:2013uba}. The RS signal is calculated using Eq.~\eqref{eq:T_RS} for all the pixels within $30\times \theta_{200c}$ of each halo center, and  it is exponentially suppressed beyond $10\times\theta_{200c}$ to avoid edges and sudden drops in the mock map.     
	
	The CMB map is simulated using CAMB\footnote{\href{https://camb.info/}{camb.info}} \citep{Lewis:1999bs}. We estimate the CMB weak lensing effect for each halo as (see \cite{Lewis:2006fu} and Eqs. 6-8 in \cite{Baxter:2014frs})
	
	\begin{equation}\label{eq:lensing}
	    \Delta T_{\rm lens}(\bm \xi) = \bm \alpha(\bm \xi) \cdot \bm \nabla T(\bm \xi).
	\end{equation}
	Here $\bm \alpha$ includes the contribution from all the lenses along the line-of-sight throughout the redshift bin. The RS, CMB and lensing maps are then combined together and smoothed with a 1 arcmin beam.  We superimpose this map with 3 different white noise realizations of 0.5, 1 and 2 $\mu$K-arcmin. These experimental setups are chosen to roughly match the characteristics of the CMB-S4 experiment \citep{CMB-S4science}. Note that CMB-S4 is expected to detect many more objects than the ones present in our simulations and the method is potentially applicable even if the objects are detected through other surveys. Therefore the final results presented here should be taken as a proof of concept and certainly represent a conservative lower limit to the actual S/N that can be attained.  
	
	Other nuisance effects such as kSZ, thermal Sunyaev Zeldovich effect (tSZ) \citep{Sunyaev1969} and the RS effect due to non-linear structure formation \citep{Seljak:1995eu,Tuluie:1995ut} are neglected in the mapmaking process. Since for spherical clusters these signals are symmetric, they will not be selected by the filter. Point sources are also neglected by the same token. %Moreover, the projection tensor $\bm q_{ij}$ eliminates any signal that does not have tangential correlations in the sky, which justifies neglecting these signals even further. 
	We also neglect rotational kSZ \citep{Chluba:2002es, Cooray:2001vy} and any internal flows \citep{Nagai:2002nw} or temperature gradients within the clusters. These effects could potentially be confused with RS and appear as spurious signal in the reconstruction of the transverse velocity for individual halos. However, the critical point is that the pairwise estimator is only sensitive to signals that are correlated with the distance between the clusters. Therefore, any effect which is not inherently a function of separation, $r$ will average out for a sufficiently large catalog. We leave a detailed study of these extra noise components for future work and only focus on extracting the signal from lensed CMB, and instrumental noise.

\section{Analysis}\label{sec:analysis}
    In order to reconstruct the transverse velocities, we  apply the DGD3 filter (Eq.~\ref{eq:DGD3_filter}) to the final simulated map from the previous section, one halo at a time. We enhance the performance of the filter by  pre-processing the map in two steps. First, as described in \S\ref{sec:filter}, we apply a Gaussian HPF with FWHM=$5'$ to suppress any large scale gradients that could be potentially picked up by the filter. Second, similar to what was done in  \cite{Maturi:2006hz}, we de-lens the map using the inverse of Eq.~\eqref{eq:lensing}. Here we are assuming that the mass and 3D location of the halo (as determined by an spectroscopic survey) are  precisely known,   and therefore errors associated with center mislocation and imperfect delensing are neglected \citep{Calafut:2017mzp}.
    
\begin{figure}[t!]
        \centering
        \includegraphics[width=\columnwidth]{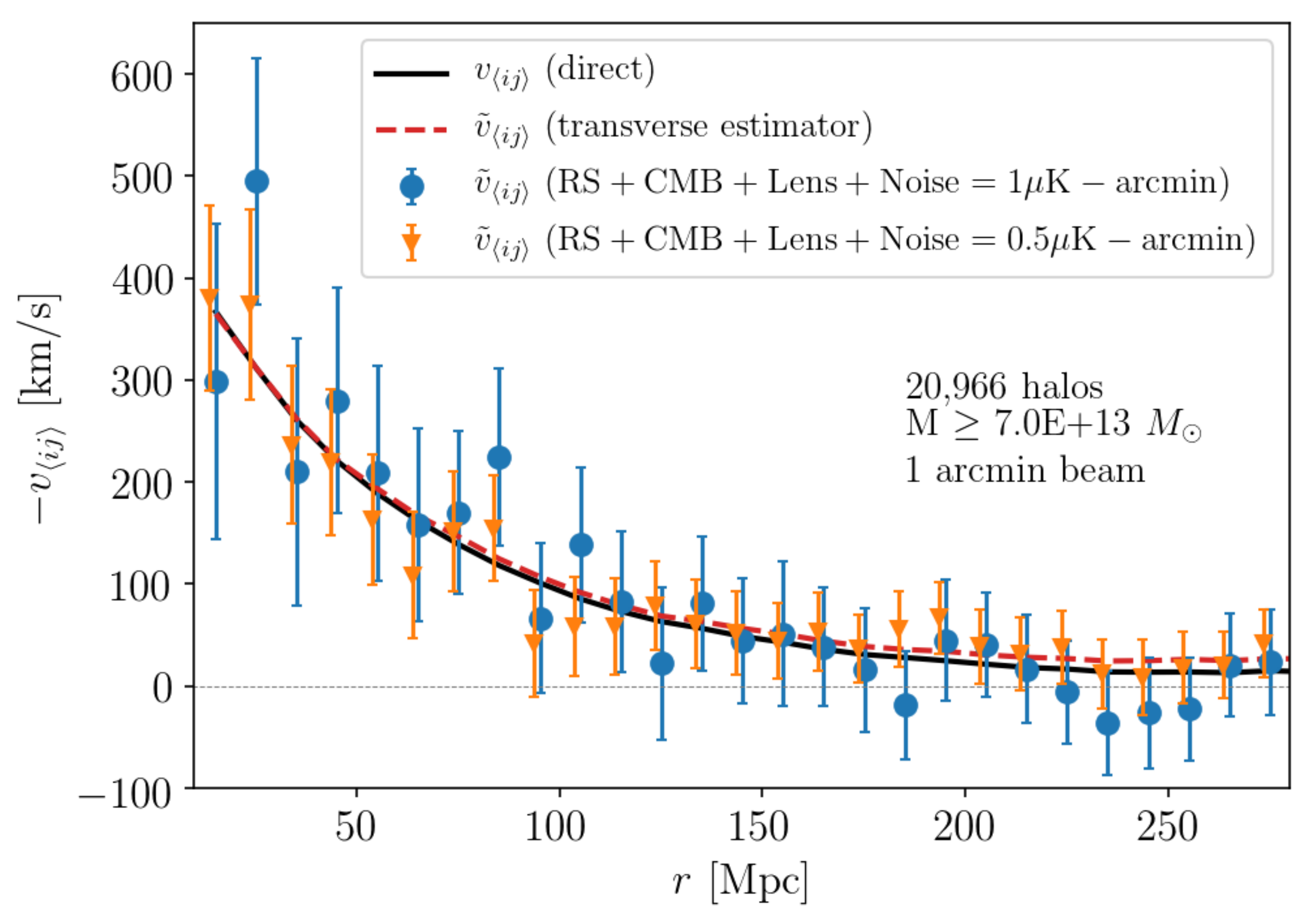}
        \caption{Mean pairwise velocities in the simulations. The black line is the mean pairwise velocity calculated directly from the 3D N-body simulation. The dashed red line represents the estimated mean pairwise velocity using the transverse components in the N-body simulation. The blue circles (orange triangles) represent the reconstructed pairwise velocities from the simulated CMB temperature map as described in \S\ref{sec:simul} with the noise level 1$\mu$K-arcmin (0.5$\mu$K-arcmin). The error bars on the black and red lines are negligible.}
        \label{fig:vij}
    \end{figure}

    We apply the DGD3 filter in the $\hat{ \bm\theta}$ and $\hat{\bm\phi}$ directions respectively, to reconstruct the velocities in these directions (see Eq.~\eqref{eq:filtered_t_i}). Next, we insert the reconstructed transverse velocity vectors in Eq.~\eqref{eq:vij_t} to obtain the pairwise velocities for all values of pairwise distances $r$, which we eventually bin with $\Delta r=10~\rm Mpc$ up to $r_{\rm max}=300~\rm Mpc$. The results are plotted for two different noise configuration in Fig.~\ref{fig:vij}.
    %The results for the 2$\mu$K-arcmin noise configuration are 
    
    The error bars are calculated using the delete-$d$ jackknife method \citep{Escoffier:2016qnf} for 100 subsamples\footnote{The average error on the elements of the covariance matrix calculated with 100 jackkinfe sub-samples is estimated to be 3.3\% (diagonal) and 2.3\% (off-diagonal). }. The S/N for each noise configuration is then calculated using
    \begin{equation}
        \widetilde{S/N} = \frac{\tilde{\bm v}^T_{\pij} C^{-1}_v \bm v_{\pij}}{\sqrt{{\bm v}^T_{\pij} C^{-1}_v \bm v_{\pij}}},
    \end{equation}
    where ${\bm v}_{\pij}$ and $\tilde{\bm v}_{\pij}$ are the binned arrays of the direct and estimator-reconstructed pairwise velocities and $C_v$ is the covariance matrix (the diagonal elements are the error bars at each $r$ in the Fig.~\ref{fig:vij}).

    Although the errors on individual velocity reconstructions from applying the DGD3 filter to the mock maps are significantly large, the  estimator still performs remarkably well in recovering the pairwise velocities. The resulting S/N is 8.5, 5.2 and 2.3 respectively for the $0.5\mu$K-arcmin, $1\mu$K-arcmin and $2\mu$K-arcmin noise configurations. For reference, the S/N for a pure RS map is about 80. Among the sources of noise that we are taking into account, the primary CMB is the most troublesome because of its inherent scale dependent correlation. The gradients in the CMB that appear as spurious signal do not average out in the estimator as efficiently as lensing and instrumental noise. Using an optimal matched filter would certainly increase the performance of the estimator and yield a higher S/N.
	
\section{Summary and Discussion}\label{sec:concl}

In this paper, we introduced an estimator for the mean pairwise velocity which is based on the measurement of the transverse components from observational data. We exploited the Rees-Sciama (RS) effect as a probe of the transverse velocities and used a heuristic filter to extract this signal from simulated maps. Although in the presence of the CMB and instrumental noise the individual velocities were not well-reconstructed, we demonstrated that by using the estimator it is possible to measure the pairwise velocities with a significant signal-to-noise ratio. When emulating a CMB-S4-like experiment with a noise level of 1 $\mu$K-arcmin on an area of $40\times40$ deg$^2$ with $\sim21000$ halos at $z=0.5$ in the mass range $M >7\times10^{13}\rm M_\odot$, we attain S/N $=5.2$ for the reconstructed mean pairwise velocity. 

The power of the estimator lies within its implicit dependence on the distance between the halos: any nuisance or spurious signal which does not depend on the distance between the clusters would statistically average out. The primary CMB temperature gradients are exempt from this property because of their inherent scale dependence. Indeed, we found that among all the noise components considered in this study, the primary CMB was the most detrimental. This, however, can potentially be alleviated by replacing our heuristic filter by an optimal matched filter \citep{Haehnelt:1995dg,Maturi:2006hz} which uses prior knowledge of the CMB and noise power spectra to minimize the spurious contribution of these components.

Even though we were able to extract the RS effect with a sub-optimal filter and reconstruct the mean pairwise velocities with a significant S/N, our result relies on several simplistic assumptions. We simulated the RS map using an NFW profile for spherically symmetric halos with known mass and angular positions and distances in the sky. Despite the fact that slight modifications to the density profile and the total mass are not expected to change the results significantly, any asymmetry in the halos or center mislocation would be propagated in the velocity reconstruction process. Effects that are azimuthally symmetric in the plane of sky such as kSZ, tSZ and intrinsic RS due to non-linear structure formation were not taken into account because they would be filtered out by the anti-symmetric filter used in this study. Additional anti-symmetric sources of noise such as the rotational kSZ, and internal temperature gradients in the clusters were also neglected, because they would statically average out in the estimator. In general, the effect of any noise source that does not depend on the distance between the clusters is expected to become less prominent as the number of halos increase. An accurate assessment of the relevance of these issues requires more detailed simulations and analyses than what has been presented here. 

The only anti-symmetric contaminant that we considered in the analysis was the weak lensing of the clusters which also creates a dipolar pattern around the halo center, aligned with the temperature gradient of the background CMB. The amplitude of this effect is orders of magnitude smaller than the CMB itself, but it can be larger than the RS signal induced by the transverse motion of clusters and hence diminish the accuracy of individual velocity reconstruction. Nevertheless, weak lensing is not a problematic noise component in our analysis because the pairwise estimator averages this contaminant out sufficiently well. Even though we assumed that almost perfect de-lensing is possible, ignoring this step does not lower the pairwise reconstruction signal-to-noise significantly. Skipping the de-lensing step for the 1 $\mu$K-arcmin and 0.5 $\mu$K-arcmin noise configurations lowers the S/N respectively from 5.2 and 8.5, to 4.5 and 8.0.

Assuming isotropy in the velocity field, the mean pairwise velocity measured with the transverse component (via the RS effect) should yield the same result as the one measured with the radial component (via kSZ). However, any deviation between the two could diagnose systematics in either measurement or imply existence of non-standard cosmology (e.g. rotational velocity fields). Combining the complimentary velocity measurements from RS and kSZ can put tight constraints on cosmological models and the statistics of the large scale structure. Here we emphasize again that pairwise measurements obtained with the RS effect do not require a knowledge of the  baryonic physics of the cluster, as it is the case for the kSZ based approach. The only required parameters are the mass profile of the cluster and its location. 

The estimator introduced here is generally applicable to all possible probes of transverse velocities. 
In this study we focused on the RS effect to measure the transverse velocities in CMB temperature maps, though future surveys could potentially also measure transverse velocities through the study of kSZ polarization \citep{Sunyaev:1980nv, Sazonov1999,Yasini:2016pby,Deutsch:2017cja,Shimon2009,Meyers:2017rtf}. The amplitude of this effect is about an order of magnitude smaller, but it is conceivable to extract this signal by cross-correlating it with the RS map. This topic will be the focus of future work.

While this paper was in the final stages of preparation for submission, a work by \cite{Hotinli:2018yyc} came to our attention. The authors use a different approach to make a statistical detection of the transverse velocities by employing the ``moving lens" power spectrum. Their results further enhance the relevance of studying transverse velocities in the near future.

\acknowledgments

We sincerely thank Arthur Kosowsky, Pavel Motloch, Jens Chluba, Sunil Golwala, and Azadeh Fattahi for helpful discussions. We are also incredibly grateful to the referee for their detailed analysis of our work and for providing countless insightful comments. Computation for the work described in this paper was supported by the University of Southern California’s Center for High-Performance Computing. EP and NM are supported by NASA grant 80NSSC18K0403. We acknowledge using the \texttt{Scipy} library \citep{Scipy} for preparation of the results presented in this work. 
%k\ bibliographystyle{apsrev4-1}	
\bibliographystyle{aasjournal.bst}	
%{\footnotesize \setlength{\bibsep}{5pt}
%	\bibliography{cosmobib}}
%\bibliography{cosmobib}

\end{document}